\documentclass[pra, superscriptaddress, twocolumn, showpacs]{revtex4}
\usepackage{epsfig}
\usepackage{amsmath}
\usepackage{graphicx}
\usepackage{bbm}
\def\be{ \begin{equation} }
\def\ee{ \end{equation} }
\def\bea{ \begin{eqnarray} }
\def\eea{ \end{eqnarray} }
\def\bse{ \begin{subequations} }
\def\ese{ \end{subequations} }

\def\half{\tfrac12}

\begin{document}
%%%%%%%%%%%%%%%%%%%%%%%%%%%%%%%%%%%%%%%%%%%%%%%%%%%%%%%%%%%%%%%%%%%%%%%%%%%%%%%%%%%%%%%%%%%%%%%%%%%%%%%%%%%%%%%%%%

\author{I. E. Linington}
\affiliation{Department of Physics, Sofia University, James Bourchier 5 blvd, 1164 Sofia,
Bulgaria}
\affiliation{Department of Physics and Astronomy, University of Sussex, Falmer, Brighton,
BN1 9QH, United Kingdom}
\author{N. V. Vitanov}
\affiliation{Department of Physics, Sofia University, James Bourchier 5 blvd, 1164 Sofia,
Bulgaria}
\affiliation{Institute of Solid State Physics, Bulgarian Academy of Sciences,
Tsarigradsko chauss\'{e}e 72, 1784 Sofia, Bulgaria}

%%%%%%%%%%%%%%%%%%%%%%%%%%%%%%%%%%%%%%%%%%%%%%%%%%%%%%%%%%%%%%%%%%%%%%%%%%%%%%%%%%%%%%%%%%%%%%%%%%%%%%
\title{Decoherence-free preparation of Dicke states of trapped ions by collective stimulated Raman adiabatic passage}
%%%%%%%%%%%%%%%%%%%%%%%%%%%%%%%%%%%%%%%%%%%%%%%%%%%%%%%%%%%%%%%%%%%%%%%%%%%%%%%%%%%%%%%%%%%%%%%%%%%%%%
%
%%%%%%%%%%%%%%%%%%%%%%%%%%%%%%%%%%%%%%%%%%%%%%%%%%%%%%%%%%%%%%%%%%%%%%%%%%%%%%%%%%%%%%%%%%%%%%%%%%%%%%
\begin{abstract}
We propose a simple technique for the generation of arbitrary-sized Dicke states in a chain of trapped ions. The method uses global addressing of the entire chain by two pairs of delayed but partially overlapping laser pulses to engineer a collective adiabatic passage along a multi-ion dark state. Our technique, which is a many-particle generalization of stimulated Raman adiabatic passage (STIRAP), is decoherence-free with respect to spontaneous emission and robust against moderate fluctuations in the experimental parameters. Furthermore, because the process is very rapid, the effects of heating are almost negligible under realistic experimental conditions. We predict that the overall fidelity of synthesis of a Dicke state involving ten ions sharing two excitations should approach $98\%$ with currently achievable experimental parameters.
\end{abstract}
%%%%%%%%%%%%%%%%%%%%%%%%%%%%%%%%%%%%%%%%%%%%%%%%%%%%%%%%%%%%%%%%%%%%%%%%%%%%%%%%%%%%%%%%%%%%%%%%%%%%%%
\maketitle
%
%%%%%%%%%%%%%%%%%%%%%%%%%%%%%%%%%%%%%%%%%%%%%%%%%%%%%%%%%%%%%%%%%%%%%%%%%%%%%%%%%%%%%%%%%%%%%%%%%%%%%%
\section{Introduction}
\label{introduction}
%%%%%%%%%%%%%%%%%%%%%%%%%%%%%%%%%%%%%%%%%%%%%%%%%%%%%%%%%%%%%%%%%%%%%%%%%%%%%%%%%%%%%%%%%%%%%%%%%%%%%%
%
A clear understanding of many-body entanglement is centrally important for the fundamental description of microscopic systems. Within the framework of quantum information science, entanglement may be viewed as a resource for the processing of information in ways not permitted by classical logic \cite{nielsen2000}. This has stimulated an intense research effort aimed at studying the properties of multipartite entanglement, and in turn, experience gained from such studies is providing fruitful insights into the complex behaviour of condensed matter systems \cite{amico2007}. 
However, the characterisation and quantification of multipartite entanglement remains an open problem; a full classification has only been achieved for small numbers of qubits \cite{dur2000, vestraete2002, lamata2007, wu2007, bai2007} and the extension to larger systems represents a formidable challenge.

The \emph{experimental generation} and investigation of multipartite entangled states is therefore a highly desirable tool for the future characterisation of many-body entanglement. In this regard, an interesting class of such states are the \emph{Dicke states}, $\vert{}W_{m}^{N}\rangle$, in which $m$ excitations are distributed evenly amongst $N$ parties in an equal coherent superposition \cite{dicke1954, mandel1995}:
\begin{align}
\left\vert{}W_{m}^{N}\right\rangle & =\frac{1}{\sqrt{C_{m}^{N}}}\widehat{\mathcal{S}}\vert{}\underbrace{1\ldots{}1}_{m}\underbrace{0\ldots0}_{N-m}\rangle.
\label{Dicke_state_def}
\end{align}
Here, the symmetrization operator, $\widehat{\mathcal{S}}$ creates an (un-normalised) equal superposition of all distinct permutations of the $N$ qubits, which number \mbox{$C_{m}^{N}\equiv{}N!/[m!(N-m)!]$}.
Dicke states exhibit genuine multi-body entanglement \cite{toth2007, devi2007} and are also robust against particle loss and measurement \cite{stockton2003, bourennane2006}. Furthermore, the entanglement contained in such states cannot be destroyed by local operations performed on any constituent particle \cite{kiesel2007}, making them interesting from the point of view of quantum communication between many participants. The notation $\vert{}W_{m}^{N}\rangle$ emphasises that the Dicke states represent a generalisation to $m$ excitations of the more ubiquitous W-states, in which a single excitation is shared evenly between $N$ sites. We note that while W-states of up to eight ions have been created experimentally in ion traps \cite{haffner2005}, the extension to higher numbers of excitations represents a major challenge, both from a theoretical and experimental perspective. This is because the dimension of the relevant Hilbert space, $\mathcal{H}$, grows very rapidly with the number of excitations involved. 

Theoretical proposals exist for the generation of Dicke states in a number of physical systems, including ensembles of neutral atoms \cite{stockton2004, mandilara2007, thiel2007}, trapped ions \cite{Dicke_bow_tie_2008,retzker2007}, quantum dots \cite{zou2003} and using linear optics \cite{kiesel2007,thiel2007}.
Ion traps are perhaps the best suited of these for the manipulation of entanglement in matter systems since they offer an unparalleled level of experimental control; the state of individual qubits can be initialised, manipulated and read out with $>99\%$ precision, and information can be stored without loss for times many orders of magnitude longer than typical gate operation times.
By using carefully chosen laser pulses, the ions' internal and motional degrees of freedom can be coupled, and allowing different ions to interact with a common vibrational mode (the `bus-mode') then creates an effective interaction between the ions through which entanglement can be generated \cite{cirac1995}.
In this manner, entangling gates have been performed with high fidelity \cite{leibfried2003,schmidt2003}, and complex multipartite entangled states have been created \cite{haffner2005, leibfried2005}.

The generation of large coherent superpositions in contemporary ion-trap quantum information science is limited by two major sources of decoherence. These are: \mbox{(i) \emph{spontaneous emission} } as the ions decay from high-lying energy states, occupied transiently during internal transitions; \mbox{(ii) \emph{motional heating}} caused by coupling between the ions and fluctuating patch potentials on the surface of the electrodes used to trap the ions; this has the effect of adding or removing vibrational quanta from the system at random.

Various techniques have been proposed with the aim of limiting the effects of spontaneous emission. Of particular note is the \mbox{STIRAP} technique for robust transfer of population between internal levels of a single particle \cite{vitanov2001} which has recently been applied to coherent population transfer in a \emph{single} ion with over $95\%$ fidelity \cite{sorensen2006}. Here, a \emph{dark state} of the system is utilised, in which higher-energy levels always remain unoccupied. In the adiabatic limit (whereby the system remains in the dark state at all times), the technique is therefore decoherence-free with respect to spontaneous emission. Recently, it has been suggested to use variants of the traditional \mbox{STIRAP} process in order to implement two-ion quantum gates \cite{pachos2002} and to generate arbitrary entangled states of a two-spin system \cite{unanyan2001}. 

Regarding the second source of decoherence mentioned above, motional heating of the ion chain by its environment appears to be unavoidable in traditional multiple-step techniques, and the degradation to the state that this causes grows with the time taken for the state preparation. Therefore the (traditional) approach of using many sequential steps, each of which addresses an individual ion, appears unsuitable for the creation of large entangled states; the Hilbert space dimension and hence the number of steps required by an individual addressing approach increases extremely rapidly with both the number of ions and the number of quanta involved. 

A promising route to limiting the number of interaction steps is to address all of the ions simultaneously using common laser pulses (a technique known as \emph{global addressing}). In this way, the time required, and hence the effects of heating, can be dramatically reduced. Several global addressing schemes have been proposed \cite{sorensen1999b,unanyan2002,unanyan2003,Dicke_bow_tie_2008} and experimentally implemented \cite{leibfried2005} for the generation of entanglement in trapped ion systems.
An alternative route to limiting the effects of heating is to look for schemes which do not require the number of motional quanta to be controlled precisely \cite{sorensen1999b, sorensen1999, leibfried2005}. In this situation, modest vibrational heating can be tolerated and spontaneous emission is then the dominant source of decoherence. 

To date, it has not been possible to find an approach to entangled state generation which integrates the advantages of insensitivity to vibrational heating alongside a robustness against spontaneous emission. Below, we present a scheme which goes some way towards this aim, by being fully immune to spontaneous emission and largely insensitive to heating effects. We propose to generate Dicke states of \emph{arbitrary size} in a chain of trapped ions using global addressing by two pairs of overlapping, counter-intuitively ordered laser pulses. With an appropriate choice of the laser parameters, it is possible to navigate through the overall Hilbert space and into the target state whilst remaining in the multi-ion dark state at all times. Because the technique presented in this article is very rapid, the destructive effects of heating are largely circumvented, while the use of a multi-ion dark state means that spontaneous emission can be avoided completely in the adiabatic limit. Consequently, the fidelity of the state preparation using our technique can be extremely high. Moreover, the implementation is appealingly simple: it requires only two pairs of sufficiently intense and suitably delayed laser pulses, rather than sophisticated sequences of great number of pulses of precise areas, as in traditional approaches.

The remainder of this article is organised as follows: In section \ref{model}, a Hamiltonian is introduced to describe a chain of $N$ ions interacting with a pair of common laser pulses, tuned on resonance with the first red motional sideband. 
Two important properties possessed by this Hamiltonian are considered in section \ref{symmetries}, and by exploiting the symmetry of the chosen system, it is shown in subsection \ref{dark_state} that the system possesses a unique multi-ion dark state which is immune to spontaneous emission. 
In section \ref{generation}, we show that, remarkably, Dicke states of arbitrary size can be generated robustly and efficiently in two steps, by adiabatic transfer along this dark-state. 
During the transfer process, the dark-state evolves through a network of increasingly complex $N$-particle entangled states. 
The net phase acquired during the state preparation is identically zero.

Section \ref{numerical} discusses the conditions required for adiabatic following of the dark state, since this is important to the success of our technique. Other sources of decoherence, such as vibrational heating and parameter fluctuations are considered in section \ref{technical}, and it is found that the method presented in the current paper is robust against these types of decoherence also. In section \ref{conclusions}, we summarise our findings.

%%%%%%%%%%%%%%%%%%%%%%%%%%%%%%%%%%%%%%%%%%%%%%%%%%%%%%%%%%%%%%%%%%%%%%%%%%%%%%%%%%%%%%%%%%%%%%%%%%%%%%
\section{Theoretical model and dark-state structure}
\label{model}
%%%%%%%%%%%%%%%%%%%%%%%%%%%%%%%%%%%%%%%%%%%%%%%%%%%%%%%%%%%%%%%%%%%%%%%%%%%%%%%%%%%%%%%%%%%%%%%%%%%%%%
%
%%%%%%%%%%%%%%%%%%%%%%%%%%%%%%%%%%%%%%%%%%%%%%%%%%%%%%%%%%%%%%%%%%%%%%%%%%%%%%%%%%%%%%%%%%%%%%%%%%%%%%
\subsection{Definition of the problem}
\label{definition}
\begin{figure}[t]
\includegraphics[width=0.55\columnwidth]{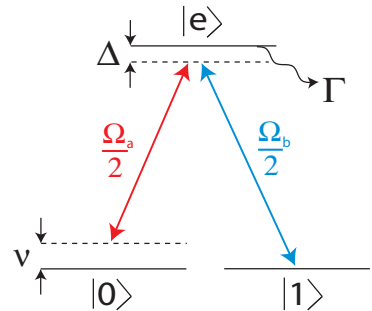}
\caption{Raman-coupled ion qubit. The computational basis states $\vert0\rangle$ and $\vert1\rangle$ are hyperfine or Zeeman sub-levels of the ion's ground state and have very good coherence properties. By contrast, decay from the excited state $\vert{}e\rangle$ is typically a significant source of decoherence in modern ion-trap quantum information processing experiments \cite{ozeri2005, aolita2007}.}
\label{three_level_ion}
\end{figure}

%%%%%%%%%%%%%%%%%%%%%%%%%%%%%%%%%%%%%%%%%%%%%%%%%%%%%%%%%%%%%%%%%%%%%%%%%%%%%%%%%%%%%%%%%%%%%%%%%%%%%%
%
We consider a chain of $N$ identical ions trapped in a linear array and cooled to their ground state. Each ion has three relevant internal levels, as shown in Fig. \ref{three_level_ion}. The computational basis states, $\vert{}0\rangle$ and $\vert{}1\rangle$, are encoded in two hyperfine or Zeeman sub-levels in the electronic ground state of the ion and typically have very good coherence properties \cite{bollinger1991, lucas2007}. Coupling between these two internal states is achieved by a two-photon Raman process, via the upper-level, $\vert{}e\rangle$ using only a single pair of pulses (labelled $a$ and $b$), each of which addresses the entire chain. Overall, the laser-frequencies, $\omega_{a}$ and $\omega_{b}$, are tuned to the first red motional sideband of the centre-of-mass mode for the transition \mbox{$\vert0\rangle\leftrightarrow\vert1\rangle$}, while $\omega_{a}$ is tuned near to the first red sideband for the \mbox{$\vert0\rangle\leftrightarrow\vert{}e\rangle$} transition, i.e.
\begin{subequations}
\begin{align}
\omega_{a} =\; & \omega_{0e}-\Delta-\nu,
\\
\omega_{b} = \;& \omega_{1e}-\Delta,
\end{align}
\end{subequations}
where $\omega_{0e}$ and $\omega_{1e}$ are the Bohr frequencies of the transitions $\vert0\rangle\leftrightarrow\vert{}e\rangle$ and $\vert1\rangle\leftrightarrow\vert{}e\rangle$ respectively, $\nu$ is the trap frequency and $\Delta$ is a constant single-photon detuning. We note that in contrast to existing Raman-coupled schemes in ion traps \cite{wineland2003}, the technique proposed below, does \emph{not} require adiabatic elimination of the state $\vert{}e\rangle$ in order to limit spontaneous emission and in fact we choose $\Delta<\nu$. In this limit [and because ultimately there will be no decay from $\vert{}e\rangle$], the sideband structure on the $\vert0\rangle\leftrightarrow\vert{}e\rangle$ transition can be resolved.
In the Lamb-Dicke limit and after making the rotating-wave approximation, the Hamiltonian for this system (with $\hbar=1$ throughout this paper) is:
\begin{align}
\tilde{H}(t) = \frac{1}{2}\sum_{j=1}^{N}\bigg\{&\Omega_{a}(t)\big[\hat{a}\vert{}e\rangle_{j}\langle0\vert_{j}\exp\left(i\Delta{}t-i\phi^{a}_{j}\right) 
\nonumber \\  
+& \hat{a}^{\dagger}\vert0\rangle_{j}\langle{}e\vert_{j}\exp\left(-i\Delta{}t+i\phi^{a}_{j}\right)\big]
\nonumber \\
+&\Omega_{b}(t)\big[\vert{}e\rangle_{j}\langle{}1\vert_{j}\exp\left(i\Delta{}t-i\phi^{b}_{j}\right)
\nonumber \\  
+&\vert{}1\rangle_{j}\langle{}e\vert_{j}\exp\left(-i\Delta{}t+i\phi^{b}_{j}\right)\big]\bigg\}.
\label{Hamiltonian}
\end{align}
Above, $\hat{a}^{\dagger}$ and $\hat{a}$ are the creation and annihilation operators for centre-of-mass phonons and $\Omega_{a}$ is an effective Rabi frequency, defined by \mbox{$\Omega_{a}\equiv\eta_{a}\tilde{\Omega}_{a}/\sqrt{N}$}, which describes the coupling between the motional and internal states of the ions \cite{james1998}. 
$\tilde{\Omega}_{a}$ and $\Omega_{b}$ are the (real-valued) bare Rabi frequencies for each laser pulse.
The single-ion Lamb-Dicke parameter for pulse $a$ is \mbox{$\eta_{a}\equiv\sqrt{\hbar{}k_{a}^{2}\cos^{2}\theta_{a}/2M\nu}$}, with $k_{a}$ being the laser
wave-number and $\theta_{a}$ the angle between the trap axis and beam direction for laser beam $a$, and $M$ the mass of one ion. Equation (\ref{Hamiltonian}) applies in the Lamb-Dicke limit, for which $\eta_{a}(n+1)^{1/2}\ll1$, where $n$ is the number of phonons in the centre-of-mass vibrational mode. Also, the trap frequency $\nu$ must satisfy 
\begin{align}
\nu\gg\left\vert\Omega_{a}(t)\right\vert,\left\vert\Omega_{b}(t)\right\vert,\Delta.
\label{james_condition}
\end{align}
Because the ions in a Paul trap are spaced unevenly along the trap
 axis, there will always be some nonzero phases
\begin{subequations}
\begin{align}
\phi_{j}^{a}\equiv{}&\phi_{L}^{a}-x_{j}k_{a}\cos\theta_{a}-\pi/2,
\\
\phi_{j}^{b}\equiv{}&\phi_{L}^{b}-x_{j}k_{b}\cos\theta_{b}-\pi/2,
\end{align}
\end{subequations}
where $x_{j}$ are the equilibrium positions of the ions and $\phi_{L}^{a,b}$ the laser phases at the trap centre at $t=0$. However, by performing a time dependent phase-transformation:
\begin{align}
\widehat{H}_{I}\left(t\right) =\mathbf{U}^{\dagger}(t)\tilde{H}_{I}\left(
t\right) \mathbf{U}(t)-i\mathbf{U}^{\dagger}(t)\frac{\partial \mathbf{U}(t)}{\partial
t}, 
\label{phase_transformation}
\end{align}
with
\begin{align}
\mathbf{U}(t) = & \exp \left[ i\sum_{j=1}^{N}\left(\Delta{}t-\phi^{a}_{j}\right) \vert{}e\rangle _{j}\langle e\vert_{j}\right]
\nonumber \\
& \;\times  \exp\left[i\sum_{j=1}^{N}(\phi^{b}_{j}-\phi^{a}_{j})\vert{}1\rangle_{j}\langle{}1\vert_{j}\right],
\end{align}
the Hamiltonian (\ref{Hamiltonian}) can be put into the following simpler form, in which each laser couples equally to all ions:
\begin{align}
\widehat{H}(t) & =\sum_{j=1}^{N}\bigg\{ \frac{\Omega_{a}(t)}{2}\big[\hat{a}\vert{}e\rangle_{j}\langle0\vert_{j}
+ \hat{a}^{\dagger}\vert0\rangle_{j}\langle{}e\vert_{j}\big]
\nonumber \\
& +\frac{\Omega_{b}(t)}{2}\big[\vert{}1\rangle_{j}\langle{}e\vert_{j}
 + \vert{}e\rangle_{j}\langle{}1\vert_{j}\big]
+\Delta\vert{}e\rangle_{j}\langle{}e\vert_{j}\bigg\}.
\label{Hamiltonian_no_phases}
\end{align}
For clarity, we shall first treat the case of equal couplings, given by (\ref{Hamiltonian_no_phases}), and choose to postpone discussion of the effects of the phase-factors, $\phi_{j}^{a,b}$ until section \ref{technical}, where several other technical considerations are also addressed.
%================================================================================================================================================================

%================================================================================================================================================================

%%%%%%%%%%%%%%%%%%%%%%%%%%%%%%%%%%%%%%%%%%%%%%%%%%%%%%%%%%%%%%%%%%%%%%%%%%%%%%%%%%%%%%%%%%%%%%%%%%%%%%
\subsection{Modelling decay from the upper state $\vert{}e\rangle$}
%%%%%%%%%%%%%%%%%%%%%%%%%%%%%%%%%%%%%%%%%%%%%%%%%%%%%%%%%%%%%%%%%%%%%%%%%%%%%%%%%%%%%%%%%%%%%%%%%%%%%%
%
While the two computational basis states can have very long coherence times [up to several minutes \cite{bollinger1991, lucas2007}], decay from the upper state $\vert{}e\rangle$ often has a significant effect on the dynamics and in current experiments can even dominate the error budget for entangling gates \cite{ozeri2005, aolita2007}. Therefore, leaving aside for the moment vibrational heating of the centre-of-mass mode [we shall return to this issue in section \ref{technical}], the overall dissipative dynamics is governed by the following master equation:
\begin{align}
\frac{\partial{}\hat{\rho}(t)}{\partial{}t} & = i\left[\hat{\rho}(t),\widehat{H}(t)\right] + \widehat{\mathcal{L}}\hat{\rho}(t).
\label{Liouville}
\end{align}
Here, $\widehat{\mathcal{L}}$ is a Liouvillian super-operator corresponding to spontaneous emission from the state $\vert{}e\rangle$ into either the computational basis states, or other levels, at rate $\Gamma$. If the upper state $\vert{}e\rangle$ is occupied, even transiently during the state preparation, the overall fidelity will decrease, due to the dissipative term, $\widehat{\mathcal{L}}\hat{\rho}(t)$. By contrast, if the state $\vert{}e\rangle$ remains unoccupied at all times, then this dissipative term has no effect on the dynamics, and equation (\ref{Liouville}) can be replaced by the Schr\"odinger equation for a pure state, $\vert{}\psi(t)\rangle$.

Below, we shall describe a method for creating Dicke states using global addressing, for which the evolution takes place along a multi-ion dark state. In the adiabatic limit, the coupling between this dark state and all other states in the Hilbert space is small enough that spontaneous emission from the level $\vert{}e\rangle$ can be neglected. 
We therefore expect spontaneous emission to have only a perturbative effect on the dynamics and so choose to work in terms of a state vector, $\vert\psi(t)\rangle$, and the Schr\"odinger equation, rather than a density operator and equation (\ref{Liouville}). The effects of emission from state $\vert{}e\rangle$ can still be included, simply by making the following replacement:
\begin{align}
\Delta\mapsto\tilde{\Delta}\equiv\Delta-i\Gamma.
\label{complex_Delta}
\end{align}
The above replacement is computationally simpler than solving the full master equation, (\ref{Liouville}), and is physically equivalent to the assumption that all spontaneous emission from $\vert{}e\rangle$ takes place into ionic levels other than $\vert{}0\rangle$ and $\vert{}1\rangle$. We note that the norm of the state-vector calculated using (\ref{complex_Delta}) is not constant, since after introducing a complex detuning, the Hamiltonian is no longer Hermitian. The effects of spontaneous emission are therefore manifest through a reduction in the norm of $\vert{}\psi(t)\rangle$. By making the replacement (\ref{complex_Delta}) we are taking a conservative approach in the sense that the fidelity calculated using this reduced-norm state vector is always less than or equal to the true fidelity calculated using (\ref{Liouville}).
%
%%%%%%%%%%%%%%%%%%%%%%%%%%%%%%%%%%%%%%%%%%%%%%%%%%%%%%%%%%%%%%%%%%%%%%%%%%%%%%%%%%%%%%%%%%%%%%%%%%%%%%
\subsection{Symmetries of the Hamiltonian and reduced state-space}
\label{symmetries}
%%%%%%%%%%%%%%%%%%%%%%%%%%%%%%%%%%%%%%%%%%%%%%%%%%%%%%%%%%%%%%%%%%%%%%%%%%%%%%%%%%%%%%%%%%%%%%%%%%%%%%
%
The above Hamiltonian, (\ref{Hamiltonian_no_phases}), possesses two important properties. The first is that  the total number of excitations, 
\begin{align}
\widehat{N} = & \hat{a}^{\dagger}\hat{a}+\sum_{j=1}^{N}\left(\vert{}1\rangle_{j}\langle1\vert_{j} + \vert{}e\rangle_{j}\langle{}e\vert_{j}\right),
\end{align}
commutes with $\widehat{H}_{I}(t)$ and is therefore a conserved quantity. If the system is prepared in a state with a specific number of quanta, $m$, then the subspace of states containing $m$ quanta is closed upon evolution under the Schr\"odinger equation.
However even when the system is restricted to the subspace containing exactly $m$ quanta, the  dimension of this subspace is still \mbox{$\sum_{\epsilon=0}^{m}\sum_{\mu=0}^{m-\epsilon}C_{m-\mu}^{N}C_{\epsilon}^{m-\mu}$}, which grows rapidly with both $m$ and $N$. 

\begin{figure}
\includegraphics[width=0.95\columnwidth]{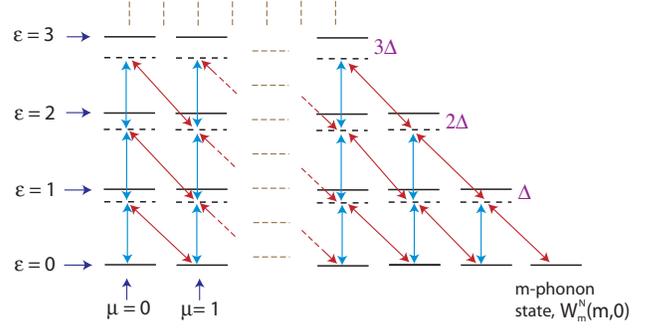}
\caption{The levels $\vert{}W_{m}^{N}(\mu,\epsilon)\rangle$ are arranged horizontally by $\epsilon$  and vertically by $\mu$. Each state is only coupled to states in neighbouring horizontal manifolds and the coupling always links states with a different value of $\epsilon$. 
As in other figures, the red couplings are proportional to $\Omega_{a}(t)$ and the blue couplings are proportional to $\Omega_{b}(t)$.}
\label{big_symm_space}
\end{figure}

A brute-force approach to solving the dynamics is clearly impractical. Fortunately however, $\widehat{H}_{I}(t)$ possesses a symmetry that permits a further simplification of the dynamics -- the Hamiltonian (\ref{Hamiltonian_no_phases}) is invariant under interchange of any two ions. Providing that the system begins in a symmetric state, it will always evolve through symmetric states at later times and with an appropriate choice of basis, the dynamics is confined to a closed $C_{2}^{m+2}$-dimensional subspace of $\mathcal{H}$ \footnote{The dimension of this subspace, i.e. $C_{2}^{m+2}$, is the \mbox{$(m+1)^{th}$} triangle-number, which is clear from Fig. \ref{big_symm_space}.}.
We choose the notation $\vert{}W_{m}^{N}(\mu,\epsilon)\rangle$ for a symmetric state of $m$ excitations shared between $N$ ions and the bus mode. $\mu$ represents the number of phonons, and $\epsilon$ is the number of ions in state $\vert{}e\rangle$. This leaves \mbox{$m-\mu-\epsilon$} ions in state $\vert{}1\rangle$ and therefore \mbox{$N-m+\mu$} ions in state $\vert{}0\rangle$. 
Formally, $\vert{}W_{m}^{N}(\mu,\epsilon)\rangle$ is defined as follows:
\begin{align}
\left\vert{}W_{m}^{N}(\mu,\epsilon)\right\rangle & = \mathcal{N}_{m}^{N}(\mu,\epsilon)\widehat{\mathcal{S}}\vert{}\underbrace{e\ldots{}e}_{\epsilon}\underbrace{1\ldots1}_{m-\mu-\epsilon}\underbrace{0\ldots0}_{N-m+\mu}\rangle\vert\mu\rangle,
\label{considered_states}
\end{align}
where $\widehat{\mathcal{S}}$ creates an (un-normalised) equal superposition of all distinct permutations of the ions' internal states, and the normalisation coefficient, $\mathcal{N}_{m}^{N}(\mu,\epsilon)$, is given by
\begin{align}
\mathcal{N}_{m}^{N}(\mu,\epsilon) = & \frac{1}{\sqrt{C_{m-\mu}^{N}C_{\epsilon}^{m-\mu}}}
\nonumber \\
= & \sqrt{\frac{\epsilon!(m-\mu-\epsilon)!(N-m+\mu)!}{N!}}.
\end{align}
All but three of the states defined by Eq. (\ref{considered_states}) are symmetric entangled states of all $N$ ions. The exceptions are the state \mbox{$\left\vert{}W_{m}^{N}(m,0)\right\rangle=\vert0_{1}\ldots0_{N}\rangle\vert{}m\rangle$} which is an $m$-phonon Fock state, with all of the ions in state $\vert{}0\rangle$, \mbox{$\left\vert{}W_{m}^{m}(0,0)\right\rangle=\vert1_{1}\ldots1_{m}\rangle\vert0\rangle$} and \mbox{$\left\vert{}W_{m}^{m}(0,m)\right\rangle=\vert{}e_{1}\ldots{}e_{m}\rangle\vert0\rangle$} which are symmetric product states of the ions' internal states, with no excitations in the bus mode. We note that $\left\vert{}W_{m}^{m}(0,0)\right\rangle$ and $\left\vert{}W_{m}^{m}(0,m)\right\rangle$ only exist if the number of excitations is the same as the number of ions, $m$, that we choose to address. We note also that in this notation, the Dicke state defined in Eq. (\ref{Dicke_state_def}) (and with zero vibrational quanta) is written \mbox{$\vert{}W_{m}^{N}\rangle\vert0\rangle\equiv\vert{}W_{m}^{N}(0,0)\rangle$}.

\begin{figure}
\includegraphics[width=0.8\columnwidth]{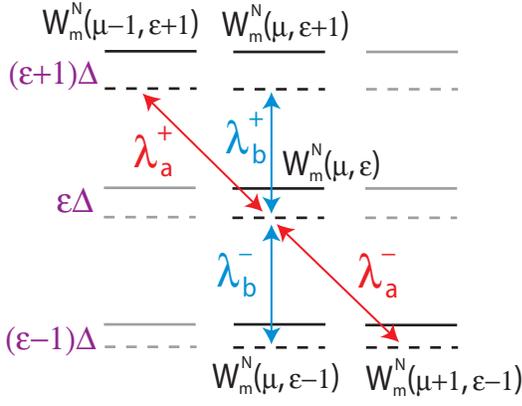}
\caption{Each symmetric state given in Eq. (\ref{considered_states}) is coupled to at most 4 other states as described by Eqs. (\ref{couplings_general}) and (\ref{lambdas}).}
\label{general_symm_coupling}
\end{figure}

The Hamiltonian (\ref{Hamiltonian_no_phases}) consists of five terms, each of which is symmetric under the interchange of any two ions and conserves the number of quanta, $\widehat{N}$. These have the following effects:
\begin{subequations}
\begin{align}
&\Big[\frac{\Omega_{a}(t)}{2}\sum_{j=1}^{N}\hat{a}\vert{}e\rangle_{j}\langle0\vert_{j}\Big]\left\vert{}W_{m}^{N}(\mu,\epsilon)\right\rangle  
\nonumber \\ & \hspace{25mm}
=\lambda_{a}^{+}(\mu,\epsilon)\left\vert{}W_{m}^{N}(\mu-1,\epsilon+1)\right\rangle, 
\\
&\Big[\frac{\Omega_{a}(t)}{2}\sum_{j=1}^{N}\hat{a}^{\dagger}\vert{}0\rangle_{j}\langle{}e\vert_{j}\Big]\left\vert{}W_{m}^{N}(\mu,\epsilon)\right\rangle   
\nonumber \\ & \hspace{25mm}
=\lambda_{a}^{-}(\mu,\epsilon)\left\vert{}W_{m}^{N}(\mu+1,\epsilon-1)\right\rangle, 
\\
&\Big[\frac{\Omega_{b}(t)}{2}\sum_{j=1}^{N}\vert{}1\rangle_{j}\langle{}e\vert_{j}\Big]\left\vert{}W_{m}^{N}(\mu,\epsilon)\right\rangle  
\nonumber \\ & \hspace{25mm} 
=\lambda_{b}^{-}(\mu,\epsilon)\left\vert{}W_{m}^{N}(\mu,\epsilon-1)\right\rangle, 
\\
&\Big[\frac{\Omega_{b}(t)}{2}\sum_{j=1}^{N}\vert{}e\rangle_{j}\langle{}1\vert_{j}\Big]\left\vert{}W_{m}^{N}(\mu,\epsilon)\right\rangle  
\nonumber \\ & \hspace{25mm}
=\lambda_{b}^{+}(\mu,\epsilon)\left\vert{}W_{m}^{N}(\mu,\epsilon+1)\right\rangle,
\\
&\Big[\Delta\sum_{j=1}^{N}\vert{}e\rangle_{j}\langle{}e\vert_{j}\Big]\left\vert{}W_{m}^{N}(\mu,\epsilon)\right\rangle  
=\epsilon\Delta\left\vert{}W_{m}^{N}(\mu,\epsilon)\right\rangle,
\end{align}
\label{couplings_general}
\end{subequations}
with the coupling matrix elements, $\lambda_{a,b}^{\pm}$, given by
\begin{subequations}
\begin{align}
\lambda_{a}^{+}(\mu,\epsilon) = & \frac{\Omega_{a}(t)}{2}\sqrt{\mu(\epsilon+1)(N-m+\mu)},
\\
\lambda_{a}^{-}(\mu,\epsilon) = & \frac{\Omega_{a}(t)}{2}\sqrt{(\mu+1)\epsilon(N-m+\mu+1)},
\\
\lambda_{b}^{-}(\mu,\epsilon) = & \frac{\Omega_{b}(t)}{2}\sqrt{\epsilon(m-\mu-\epsilon+1)},
\\
\lambda_{b}^{+}(\mu,\epsilon) = & \frac{\Omega_{b}(t)}{2}\sqrt{(\epsilon+1)(m-\mu-\epsilon)}.
\end{align}
\label{lambdas}
\end{subequations}
The above coupling structure is sketched schematically in Fig. \ref{general_symm_coupling}.
 By symmetry, it is clear that these couplings must satisfy $\lambda_{a}^{+}(\mu,\epsilon)=\lambda_{a}^{-}(\mu-1,\epsilon+1)$ and $\lambda_{b}^{+}(\mu,\epsilon)=\lambda_{b}^{-}(\mu,\epsilon+1)$. Also, from considering the boundaries of this level scheme (which only involves a finite number of states), we have $\lambda_{a}^{+}(0,\epsilon)=\lambda_{a}^{-}(\mu,0)=\lambda_{b}^{-}(\mu,0)=\lambda_{b}^{+}(\mu,m-\mu)=0$. All of these conditions are satisfied by Eqs. (\ref{lambdas}).

%================================================================================================================================================================
%================================================================================================================================================================
%
%%%%%%%%%%%%%%%%%%%%%%%%%%%%%%%%%%%%%%%%%%%%%%%%%%%%%%%%%%%%%%%%%%%%%%%%%%%%%%%%%%%%%%%%%%%%%%%%%%%%%%
\subsection{Deriving the dark-state, $\vert\varphi_{0}(t)\rangle$}
\label{dark_state}
%%%%%%%%%%%%%%%%%%%%%%%%%%%%%%%%%%%%%%%%%%%%%%%%%%%%%%%%%%%%%%%%%%%%%%%%%%%%%%%%%%%%%%%%%%%%%%%%%%%%%%
%
By definition, a dark state is an eigenstate of the Hamiltonian which does not decay by spontaneous emission. Clearly, such an eigenstate cannot involve the state $\vert{}e\rangle$ in any of the ions and therefore in the interaction-picture defined by $\widehat{H}_{I}(t)$, a dark state has zero eigenvalue, i.e.
\begin{align}
\sum_{k}H_{jk}(t)c_{k}(t) & = 0\hspace{5mm}\forall{}\;j.
\label{dark_state_condition}
\end{align}
Here $c_{k}(t)$ are the elements of the dark-state vector,
\begin{align}
\vert\varphi_{0}(t)\rangle=\sum_{k}c_{k}(t)\vert{}k\rangle,
\end{align}
and the index $k$ runs over all states defined in equation (\ref{considered_states}). The matrix elements $H_{jk}(t)$ are defined by \mbox{$H_{jk}(t)\equiv\langle{}j\vert\widehat{H}_{I}(t)\vert{}k\rangle$}.
It is certainly not obvious a priori that there is such a state satisfying (\ref{dark_state_condition}). However, we shall prove that there is one, by first assuming its existence and then calculating explicitly the elements $c_{k}(t)$.

%
%================================================================================================================================================================
%================================================================================================================================================================

To facilitate the derivation of $\vert\varphi_{0}(t)\rangle$, 
we choose to group the states $\vert{}W_{m}^{N}(\mu,\epsilon)\rangle$ into horizontal manifolds according to $\epsilon$, i.e. the number of ions in state $\vert{}e\rangle$. The other ions are found in either state $\vert{}0\rangle$ or $\vert{}1\rangle$, and all distinct permutations of the ions' states with an appropriate number of excitations are to be included. The reason that we have chosen to express the level scheme in this way is as follows. By definition, a dark state can only involve the states in the lowest ($\epsilon=0$) manifold, since all other states can decay by spontaneous emission. Therefore only this manifold and the states which are directly coupled to this manifold are relevant. We thus define the set of relevant states, $\mathcal{M}$, as the set of all states of the system with either zero or one ion in the state $\vert{}e\rangle$;
i.e. all states \mbox{$\vert{}W_{m}^{N}(\mu,0)\rangle, (\mu=0,\ldots,m)$} and \mbox{$ \vert{}W_{m}^{N}(\mu,1)\rangle, (\mu=0,\ldots,m-1)$}.

The reason that these are the `relevant' states is that the other manifolds are all (i) unoccupied and (ii) only coupled to unoccupied manifolds. Therefore, we automatically have:
\begin{subequations}
\begin{align}
\sum_{k}H_{jk}(t)c_{k}(t) = 0\hspace{5mm}j\notin\mathcal{M},
\\
\sum_{k\notin\mathcal{M}}H_{jk}(t)c_{k}(t) = 0\hspace{5mm}j\in\mathcal{M},
\end{align} 
\end{subequations}
and so Eq. (\ref{dark_state_condition}) reduces to the much simpler condition
\begin{align}
\sum_{k\in\mathcal{M}}H_{jk}(t)c_{k}(t) & = 0\hspace{5mm}j\in\mathcal{M}.
\label{simple_dark_state_condition}
\end{align}

\begin{figure}
\includegraphics[width=0.95\columnwidth]{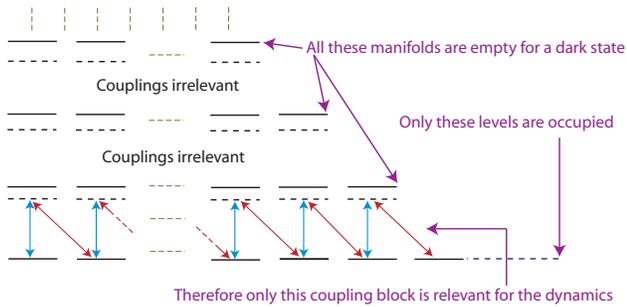}
\caption{Couplings within the relevant subspace $\mathcal{M}$. The dark state is composed solely from states in the lowest ($\epsilon=0$) manifold.}
\label{destroy_most_couplings}
\end{figure}

Therefore, if a dark state exists, the couplings to and between states that lie outside of $\mathcal{M}$ are irrelevant, and can be neglected, as shown in Fig. \ref{destroy_most_couplings}.
Hence, if a dark-state exists within the relevant subspace, $\mathcal{M}$, then this must also be a dark-state of the overall Hamiltonian given in Eq. (\ref{Hamiltonian}).
%
%================================================================================================================================================================
%================================================================================================================================================================

Upon consideration of the couplings shown in Fig. \ref{destroy_most_couplings}, it becomes clear that the relevant subspace does indeed possess a dark state, since the states of $\mathcal{M}$ are arranged in the configuration found in chain-STIRAP \cite{vitanov2001}. In particular:
\begin{itemize}
\item
There are always an odd number of states in $\mathcal{M}$; $m$ of these form an upper manifold, with detuning $\tilde{\Delta}$, while $(m+1)$ states comprise a lower-manifold, with zero detuning.
\item
Coupling between the levels alternates in a chain between couplings proportional to $\Omega_{b}(t)$ and couplings proportional to $\Omega_{a}(t)$.
\end{itemize}
For ease of notation, the abbreviations $\lambda_{a,\mu}\equiv\lambda_{a}^{+}(\mu,0)$ and $\lambda_{b,\mu}\equiv\lambda_{b}^{+}(\mu,0)$ are used from this point onwards, since within the relevant subspace, both the value of $\epsilon$ and also the label $\pm$ can be omitted without ambiguity. Furthermore, after noting that the only states that are ever involved in the dark state are Dicke states, we may revert to the simpler notation introduced in Eq. (\ref{Dicke_state_def}) and write $\vert{}W_{m-\mu}^{N}\rangle\vert\mu\rangle$ in place of $\vert{}W_{m}^{N}(\mu,\epsilon)\rangle$. The first ket denotes the internal Dicke state of $N$ ions sharing $m-\mu$ quanta and the second ket represents a motional Fock state of $\mu$ phonons. Written in block-matrix form, the dark-state condition (\ref{simple_dark_state_condition}) takes the form
\begin{align}
\begin{pmatrix}
0 & \lambda_{b,0} & 0 & \hdots & 0 & 0 & 0\\
\lambda_{b,0} & \tilde{\Delta} & \lambda_{a,1} & \hdots & 0 & 0 & 0\\
0 & \lambda_{a,1} & 0 &  \hdots & 0 & 0 & 0\\
\vdots & \vdots & \vdots &  \ddots & \vdots & \vdots & \vdots \\
0 & 0 & 0 & \hdots & 0 & \lambda_{b,m-1} & 0 \\
0 & 0 & 0 & \hdots  & \lambda_{b,m-1} & \tilde{\Delta} & \lambda_{a,m} \\
0 & 0 & 0 & \hdots & 0 & \lambda_{a,m} & 0
\end{pmatrix}
\begin{pmatrix}
c_{0} \\
0 \\
c_{1} \\
\vdots \\
c_{m-1} \\
0 \\
c_{m}
\end{pmatrix}
= & 0,
\label{matrix_form}
\end{align}
which is satisfied if 
\begin{align}
c_{\mu}\lambda_{b,\mu} + c_{\mu+1}\lambda_{a,\mu+1} = 0 & \hspace{5mm} \forall\;\; \mu=0,\ldots, m-1.
\label{recurrence}
\end{align}
We are now in a position to write down the dark state of (\ref{Hamiltonian_no_phases}):
\begin{align}
\vert{}\varphi_{0}(t)\rangle & = \sum_{\mu=0}^{m}c_{\mu}(t)\vert{}W_{m-\mu}^{N}\rangle\vert\mu\rangle,
\label{dark_state_def}
\end{align}
where the amplitudes $c_{\mu}(t)$ satisfy
\begin{align}
c_{\mu}(t) = &
 \frac{(-1)^{\mu}}{\Lambda}\prod_{k=0}^{\mu-1}\lambda_{b,k}(t)\prod_{l=\mu+1}^{m}\lambda_{a,l}(t)
\label{dark_coeffs}
\end{align}
and the normalisation factor $\Lambda$ is given by:
\begin{align}
\Lambda & = \sqrt{\sum_{\mu=0}^{m}\left\{\prod_{k=0}^{\mu-1}\lambda_{b,k}^{2}\prod_{l=\mu+1}^{m}\lambda_{a,l}^{2}\right\}}.
\label{dark_norm}
\end{align}
We note that in Eqs. (\ref{dark_coeffs}) and (\ref{dark_norm}), the convention
\begin{align}
\prod_{k=r}^{s<r}f_{k}=1
\end{align}
has been used.
The state $\vert{}\varphi_{0}(t)\rangle$ is an eigenstate of the total Hamiltonian, (\ref{Hamiltonian_no_phases}) with eigenvalue zero. Moreover, it can immediately be seen that the dark subspace of $\widehat{H}(t)$ is non-degenerate. Indeed, the recurrence relation (\ref{recurrence}) together with the normalisation condition, \mbox{$\langle\varphi_{0}(t)\vert\varphi_{0}(t)\rangle=1$}, defines $\vert\varphi_{0}(t)\rangle$ \emph{uniquely}, up to an arbitrary phase, so there is no other eigenstate of $\widehat{H}(t)$ with zero eigenvalue.

%%%%%%%%%%%%%%%%%%%%%%%%%%%%%%%%%%%%%%%%%%%%%%%%%%%%%%%%%%%%%%%%%%%%%%%%%%%%%%%%%%%%%%%%%%%%%%%%%%%%%%
\section{Generation of Dicke states}
\label{generation}
%%%%%%%%%%%%%%%%%%%%%%%%%%%%%%%%%%%%%%%%%%%%%%%%%%%%%%%%%%%%%%%%%%%%%%%%%%%%%%%%%%%%%%%%%%%%%%%%%%%%%%
%
As shown above, the composition of the dark state depends upon the ratio of the Rabi frequencies $\Omega_{a}(t)$ and $\Omega_{b}(t)$. It is crucial for the technique proposed here, $\vert\varphi_{0}(t)\rangle$ can be transformed smoothly between a product state and an entangled Dicke state, simply by a controlled manipulation of the Rabi frequencies. Therefore, by enforcing adiabatic evolution, the entire state preparation can be performed without the system ever leaving the dark state.
This process is detailed in the following three subsections
%
%%%%%%%%%%%%%%%%%%%%%%%%%%%%%%%%%%%%%%%%%%%%%%%%%%%%%%%%%%%%%%%%%%%%%%%%%%%%%%%%%%%%%%%%%%%%%%%%%%%%%%
\subsection{Properties of the dark-state}
%%%%%%%%%%%%%%%%%%%%%%%%%%%%%%%%%%%%%%%%%%%%%%%%%%%%%%%%%%%%%%%%%%%%%%%%%%%%%%%%%%%%%%%%%%%%%%%%%%%%%%
%
The properties of the state $\vert\varphi_{0}\rangle$ that we will use are:
\begin{subequations}
\begin{align}
\vert\Omega_{a}\vert\gg\vert\Omega_{b}\vert\longrightarrow\vert{}c_{\mu+1}\vert\ll\vert{}c_{\mu}\vert\hspace{5mm}(\mu=0,\ldots,m-1),
\nonumber\\
\vert\Omega_{a}\vert\ll\vert\Omega_{b}\vert\longrightarrow\vert{}c_{\mu+1}\vert\gg\vert{}c_{\mu}\vert\hspace{5mm}(\mu=0,\ldots,m-1),
\nonumber
\end{align}
\end{subequations}
and in particular:
\begin{subequations}
\begin{align}
&\lim_{\vert\Omega_{b}\vert/\vert\Omega_{a}\vert\rightarrow0}\left\{\vert{}\varphi_{0}\rangle\right\}\longrightarrow\vert{}W_{m}^{N}\rangle\vert0\rangle,
\label{limit1}
\\
&\lim_{\vert\Omega_{a}\vert/\vert\Omega_{b}\vert\rightarrow0}\left\{\vert{}\varphi_{0}\rangle\right\}\longrightarrow\vert{}0\ldots0\rangle\vert{}m\rangle.
\label{limit2}
\end{align}
\end{subequations}
Equations (\ref{limit1}) and (\ref{limit2}) are exactly the properties of the dark state used to perform complete population transfer in STIRAP and chain-STIRAP, and which have been studied extensively \cite{vitanov2001, vitanov1998, vitanov1997b}. Here, however, the STIRAP approach is extended to create entanglement in a multi-particle system, rather than to transfer population between different levels corresponding to a single particle. Drawing on the experience gained from previous studies in other \mbox{STIRAP}-type systems, we choose to use overlapping Gaussian pulse shapes, as these result in a favourable convergence to the adiabatic limit with increasing pulse area
\begin{subequations}
\begin{align}
\Omega_{a}(t) & = \Omega_{0}\exp\left(\frac{-(t+\tau)^{2}}{T^{2}}\right),
\\
\Omega_{b}(t) & = \Omega_{0}\exp\left(\frac{-(t-\tau)^{2}}{T^{2}}\right).
\end{align}
\label{pulses}
\end{subequations}
We define the terms forward \mbox{STIRAP} and reverse \mbox{STIRAP} as follows:

\begin{itemize}
\item
Forward STIRAP ($\tau>0$): When pulse $a$ precedes pulse $b$, we have:
\begin{align}
\lim_{t\rightarrow-\infty}\left\{\frac{\Omega_{b}}{\Omega_{a}}\right\}=0\hspace{6mm}\lim_{t\rightarrow\infty}\left\{\frac{\Omega_{a}}{\Omega_{b}}\right\}=0,
\end{align}
and so an initial state $\vert{}W_{m}^{N}\rangle\vert0\rangle$ will be completely transferred into the state $\vert0\ldots0\rangle\vert{}m\rangle$.
In the terminology of single-particle \mbox{STIRAP}, $\Omega_{a}$ is the Stokes pulse and $\Omega_{b}$ is the pump pulse.
\item
Reverse STIRAP ($\tau<0$): When pulse $b$ precedes pulse $a$, we have:
\begin{align}
\lim_{t\rightarrow-\infty}\left\{\frac{\Omega_{a}}{\Omega_{b}}\right\}=0\hspace{6mm}\lim_{t\rightarrow\infty}\left\{\frac{\Omega_{b}}{\Omega_{a}}\right\}=0,
\end{align}
and $\vert\varphi_{0}(t)\rangle$ evolves from $\vert0\ldots0\rangle\vert{}m\rangle$ before the pulses, to $\vert{}W_{m}^{N}\rangle\vert{}0\rangle$ afterwards.
Here the roles of the two pulses are reversed: $\Omega_{b}$ is now the Stokes pulse whereas $\Omega_{a}$ is now the pump pulse.
\end{itemize}
We note that both forward and reverse \mbox{STIRAP} utilise a counter-intuitive pulse ordering; the difference between them is simply the direction of the adiabatic transfer. 
%
%%%%%%%%%%%%%%%%%%%%%%%%%%%%%%%%%%%%%%%%%%%%%%%%%%%%%%%%%%%%%%%%%%%%%%%%%%%%%%%%%%%%%%%%%%%%%%%%%%%%%%
\subsection{State preparation by adiabatic passage}
\label{preparation}
%%%%%%%%%%%%%%%%%%%%%%%%%%%%%%%%%%%%%%%%%%%%%%%%%%%%%%%%%%%%%%%%%%%%%%%%%%%%%%%%%%%%%%%%%%%%%%%%%%%%%%
%
\noindent
Dicke states of any number of ions and quanta can be prepared by applying just two pairs of laser pulses, as follows:
\begin{enumerate}
\item
The chain is cooled to its vibrational ground state, $\vert0\rangle$, with $m$ of the ions initialised in state $\vert{}1\rangle$ and the others in state $\vert0\rangle$.
\item
Simultaneously addressing these $m$ ions on the red-sideband, but leaving the other $N-m$ ions in their initial state, the pulses given by Eq. (\ref{pulses}) are applied, with $\tau>0$ (i.e. forward \mbox{STIRAP}). This has the following effect:
\begin{align}
\vert\underbrace{1\ldots1}_{m}\underbrace{0\ldots0}_{N-m}\rangle\vert0\rangle\longrightarrow\vert\underbrace{0\ldots00\ldots0}_{N}\rangle\vert{}m\rangle.
\label{stage_1}
\end{align}
If it is technically demanding to address these $m$ ions simultaneously without addressing the other $N-m$ ions in the chain, this Fock state preparation may also be achieved in a sequential manner, by performing stage 2 repeatedly on a single ion.
\item
\emph{All N} of the ions are addressed simultaneously by applying the pulses of Eq. (\ref{pulses}) again, but this time using reverse \mbox{STIRAP} (i.e. $\tau<0$). This guides the dark state of the system into a Dicke-symmetric state, as required
\begin{align}
\vert0\ldots0\rangle\vert{}m\rangle\longrightarrow\vert{}W_{m}^{N}\rangle\vert0\rangle.
\label{stage_2}
\end{align}
\end{enumerate}
As an aside, we note that at the end of stage 2, the system resides in a vibrational number state. A detailed discussion of the generation of motional Fock states by global addressing can be found in \cite{Fock_2008}, wherein a comparison of the heating effects and adiabaticity requirements for Fock state generation by single- and multiple-ion addressing is also given.
%
%%%%%%%%%%%%%%%%%%%%%%%%%%%%%%%%%%%%%%%%%%%%%%%%%%%%%%%%%%%%%%%%%%%%%%%%%%%%%%%%%%%%%%%%%%%%%%%%%%%%%%
\subsection{Example of Dicke state preparation}
\label{example_case}
\begin{figure}
\includegraphics[width=0.95\columnwidth]{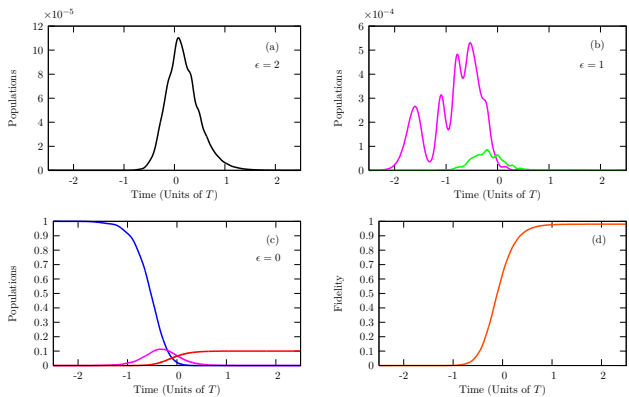}
\caption{Dicke state preparation for the case $N=5$, $m=2$. Different curves within frames (a), (b) and (c) correspond to different values of $\mu$. Frame (d) shows the time-dependent overlap with the target Dicke state. Parameters are \mbox{$\Omega_{0}T=50$}, \mbox{$\Gamma{}T=2$}, $\Delta{}T = 0$ and $\tau=-0.6T$.}
\label{dark_fig_all}
\end{figure}

%%%%%%%%%%%%%%%%%%%%%%%%%%%%%%%%%%%%%%%%%%%%%%%%%%%%%%%%%%%%%%%%%%%%%%%%%%%%%%%%%%%%%%%%%%%%%%%%%%%%%%
%
We have tested the technique described above by numerical solution of the Schr\"odinger equation for up to 10 ions and sample results are shown in Fig. \ref{dark_fig_all}. Note that the two adiabatic passage steps in the Dicke state preparation differ only in the number of ions that are addressed [$m$ in the first step and $N$ (\mbox{$N>m$}) in the second]. Therefore, for simplicity, only the second stage is plotted, starting in $\vert0\ldots0\rangle\vert{}m\rangle$ and ending in $\vert{}W_{m}^{N}\rangle\vert0\rangle$. Figure \ref{dark_fig_all} shows the example case of $N=5$, $m=2$, for which the total number of coupled many-particle states is $\dim\mathcal{H}=51$. These are grouped in three frames according to $\epsilon=0,1,2$, and then according to $\mu$ within each frame. States with the same $\epsilon$ and $\mu$ are equivalent up to a permutation of the ions and so for clarity only one curve is shown for each value of $\mu$ and $\epsilon$. Therefore, the population of only six out of the overall 51 basis states is plotted.
We draw particular attention to the red curve in Fig. \ref{dark_fig_all}c, which shows the population of $\vert11000\rangle\vert0\rangle$. This state, along with 9 other basis states with the positions of the ions permuted, comprise the Dicke state, $\vert{}W_{2}^{5}\rangle\vert0\rangle$, which explains the final state population of $\sim0.1$ for the red curve in Fig. \ref{dark_fig_all}c.

In the example shown, around $98.5\%$ of the population remains in the dark-subspace (frame(c)) during the transfer process. The slight leakage of around $1.5\%$ into decaying states is due to non-adiabatic couplings. The conditions for adiabaticity are discussed in the following section.
%================================================================================================================================================================
%================================================================================================================================================================

%%%%%%%%%%%%%%%%%%%%%%%%%%%%%%%%%%%%%%%%%%%%%%%%%%%%%%%%%%%%%%%%%%%%%%%%%%%%%%%%%%%%%%%%%%%%%%%%%%%%%%
\section{Adiabaticity requirements}
\label{numerical}
%%%%%%%%%%%%%%%%%%%%%%%%%%%%%%%%%%%%%%%%%%%%%%%%%%%%%%%%%%%%%%%%%%%%%%%%%%%%%%%%%%%%%%%%%%%%%%%%%%%%%%
%
%%%%%%%%%%%%%%%%%%%%%%%%%%%%%%%%%%%%%%%%%%%%%%%%%%%%%%%%%%%%%%%%%%%%%%%%%%%%%%%%%%%%%%%%%%%%%%%%%%%%%%
\subsection{Definition of the adiabatic basis}
\label{adiabatic_basis}
%%%%%%%%%%%%%%%%%%%%%%%%%%%%%%%%%%%%%%%%%%%%%%%%%%%%%%%%%%%%%%%%%%%%%%%%%%%%%%%%%%%%%%%%%%%%%%%%%%%%%%
%
%%%%%%%%%%%%%%%%%%%%%%%%%%%%%%%%%%%%%%%%%%%%%%%%%%%%%%%%%%%%%%%%%%%%%%%%%%%%%%%%%%%%%%%%%%%%%%%%%%%%%%
\subsubsection{Adiabatic states}
%%%%%%%%%%%%%%%%%%%%%%%%%%%%%%%%%%%%%%%%%%%%%%%%%%%%%%%%%%%%%%%%%%%%%%%%%%%%%%%%%%%%%%%%%%%%%%%%%%%%%%
%
The term \emph{adiabatic basis} refers to the time-dependent basis formed by the instantaneous eigenstates of the Hamiltonian, (\ref{Hamiltonian_no_phases}).
So far in this paper, we have concentrated on only one of the adiabatic states, namely $\vert\varphi_{0}(t)\rangle$, which has eigenvalue zero. Below, it will prove useful to consider the other eigenstates, $\vert{}\varphi(t)\rangle$, and eigenvalues, $E_{\varphi}(t)$. In the adiabatic basis, the Hamiltonian can be written as
\begin{align}
\widehat{H}_{I}(t) & = \sum_{\varphi}E_{\varphi}(t)\vert\varphi(t)\rangle\langle\varphi(t)\vert
\end{align}
and a general state-vector can be expressed as follows:
\begin{align}
\vert\psi(t)\rangle & = \sum_{\varphi}a_{\varphi}(t)\exp\left(-i\int_{t_{i}}^{t}E_{\varphi}(t')\;dt'\right)\vert\varphi(t)\rangle.
\label{adiabatic_state_vector}
\end{align}
We note that for non-zero $\Gamma$, the energies $E_{\varphi}(t)$ can be complex (with negative imaginary component). This makes clear physical sense, since the only non-decaying eigenstate of $\widehat{H}_{I}(t)$ in this situation is the dark state, $\vert\varphi_{0}(t)\rangle$.

%%%%%%%%%%%%%%%%%%%%%%%%%%%%%%%%%%%%%%%%%%%%%%%%%%%%%%%%%%%%%%%%%%%%%%%%%%%%%%%%%%%%%%%%%%%%%%%%%%%%%%
\subsubsection{Non-adiabatic couplings}
%%%%%%%%%%%%%%%%%%%%%%%%%%%%%%%%%%%%%%%%%%%%%%%%%%%%%%%%%%%%%%%%%%%%%%%%%%%%%%%%%%%%%%%%%%%%%%%%%%%%%%
%
For a stationary system [with time-independent Hamiltonian], the coefficients, $a_{\varphi}(t)$ in Eq. (\ref{adiabatic_state_vector}) would be constants. However, the explicit time dependence of Eq. (\ref{Hamiltonian_no_phases}) introduces coupling between the different eigenstates, $\vert\varphi(t)\rangle$:
\begin{align}
\frac{\partial{}a_{\varphi}(t)}{\partial{}t} & = -\sum_{\varphi'}\big\langle\varphi(t)\big\vert\frac{\partial}{\partial{}t}\big\vert\varphi'(t)\big\rangle{}a_{\varphi'}(t)
\nonumber \\
&\hspace{5mm}\times\exp\left(i\int_{t_{i}}^{t}\left[E_{\varphi}(t')-E_{\varphi'}(t')\right]\;dt'\right).
\label{nonadiabatic_couplings}
\end{align}
In the \emph{adiabatic limit}, the splittings between the various $E_{\varphi}(t)$ are large enough and the variation of $\widehat{H}_{I}(t)$ is slow enough that these couplings have a negligible effect, and a system prepared in an adiabatic state remains in that state without making transitions. The only effect on the state is that it acquires a dynamical phase, $\int_{t_{i}}^{t_{f}}E_{\varphi}(t')\;dt'$, which for the dark-state, $\vert\varphi_{0}(t)\rangle$, is identically zero.

By contrast, if the splitting between the adiabatic energies is too small, then the couplings given in (\ref{nonadiabatic_couplings}) can cause transitions between the eigenstates of $\widehat{H}_{I}(t)$. The terms on the right-hand side of Eq. (\ref{nonadiabatic_couplings}) are therefore referred to as \emph{non-adiabatic couplings}. Such couplings are harmful for the technique proposed above, since they can cause a leakage of population away from the desired state and a concomitant reduction in the fidelity. It is therefore very important to know the conditions necessary in order to reach the adiabatic limit -- these are studied in the remainder of this section.

%%%%%%%%%%%%%%%%%%%%%%%%%%%%%%%%%%%%%%%%%%%%%%%%%%%%%%%%%%%%%%%%%%%%%%%%%%%%%%%%%%%%%%%%%%%%%%%%%%%%%%
\subsection{Adiabaticity conditions}
\label{adiabaticity}
%%%%%%%%%%%%%%%%%%%%%%%%%%%%%%%%%%%%%%%%%%%%%%%%%%%%%%%%%%%%%%%%%%%%%%%%%%%%%%%%%%%%%%%%%%%%%%%%%%%%%%
%
%%%%%%%%%%%%%%%%%%%%%%%%%%%%%%%%%%%%%%%%%%%%%%%%%%%%%%%%%%%%%%%%%%%%%%%%%%%%%%%%%%%%%%%%%%%%%%%%%%%%%%
\subsubsection{Dependence on $\Delta$ and $\Omega_{0}$}
%%%%%%%%%%%%%%%%%%%%%%%%%%%%%%%%%%%%%%%%%%%%%%%%%%%%%%%%%%%%%%%%%%%%%%%%%%%%%%%%%%%%%%%%%%%%%%%%%%%%%%
%
In order to remain in the state $\vert\varphi_{0}(t)\rangle$ throughout the entire process, the energy splitting in the eigenspectrum of $\widehat{H}_{I}(t)$ must always be large enough that the effects of non-adiabatic couplings between $\vert\varphi_{0}(t)\rangle$ and all other eigenstates of $\widehat{H}_{I}(t)$ are negligible. Since we are interested in the adiabatic limit, a perturbative treatment of the non-adiabatic couplings is valid, and by this reasoning it is clear that the most significant non-adiabatic couplings are to those states whose instantaneous energies are closest to the dark state energy, i.e. zero. We choose the label $E_{1}(t)$ for the eigenvalue of $\widehat{H}_{I}(t)$ that is closest to zero, and $\vert\varphi_{1}(t)\rangle$ for the corresponding eigenstate.

An analytic derivation of the energy $E_{1}$ appears to be highly demanding. Fortunately however, for the purposes of estimating the adiabaticity requirements of the above process it is sufficient to determine the functional dependence of $E_{1}$ on the Rabi frequency, $\Omega_{0}$ and the detuning, $\Delta$.
In order to do this, we note that $E_{1}$ is a root of the characteristic polynomial:
\begin{align}
M_{2m+1} = 0,
\label{characteristic_polynomial}
\end{align}
where the determinant $M_{2m+1}$ is defined by
\begin{small}
\begin{align}
M_{2m+1} = \left\vert
\begin{array}{ccccccc}
-E_{\varphi} & \lambda_{b,0} & 0 & \hdots & 0 & 0 & 0\\
\lambda_{b,0} & \Delta-E_{\varphi} & \lambda_{a,1} & \hdots & 0 & 0 & 0\\
0 & \lambda_{a,1} & -E_{\varphi} &  \hdots & 0 & 0 & 0\\
\vdots & \vdots & \vdots &  \ddots & \vdots & \vdots & \vdots \\
0 & 0 & 0 & \hdots & -E_{\varphi} & \lambda_{b,m-1} & 0 \\
0 & 0 & 0 & \hdots  & \lambda_{b,m-1} & \Delta-E_{\varphi} & \lambda_{a,m} \\
0 & 0 & 0 & \hdots & 0 & \lambda_{a,m} & -E_{\varphi}
\end{array}\right\vert.
\label{odd_determinant}
\end{align}
\end{small}
In appendix \ref{appendix}, it is shown that the solutions of Eq. (\ref{characteristic_polynomial}) for which $E_{\varphi}\neq0$ can be expressed as $P(z)=0$, where $P(z)$ is an $m^{th}$-order polynomial in the dimensionless parameter \mbox{$z\equiv{}E_{\varphi}(E_{\varphi}-\Delta)/\Omega_{0}^{2}$}. The coefficients in $P(z)$ depend on $m$, $N$ and the time, $t$, but importantly, they are independent of both $\Omega_{0}$ and $\Delta$. We are interested in a particular root of $P(z)$, corresponding to the energy $E_{1}$, and choose to write this root as $z=\gamma_{m}^{N}(t)$, i.e.
\begin{align}
E_{1}(t) = \frac{1}{2}\left(\Delta-\sqrt{\Delta^{2}+4\Omega_{0}^{2}\gamma_{M}^{N}(t)}\right).
\end{align}
Now, the larger the value of $E_{1}(t)$ at any given time, the smaller the effect of non-adiabatic couplings.
Therefore, just as in three-state \mbox{STIRAP} \cite{vitanov2001} and other chain-\mbox{STIRAP} models \cite{vitanov1998}, adiabaticity can be improved either by increasing the pulse amplitude $\Omega_{0}$ or reducing the single-photon detuning $\Delta$. It is interesting to consider two specific limits:
\begin{itemize}
\item
\emph{Zero single-photon detuning}: In the case $\Delta=0$, the adiabatic energies come in sign-conjugate pairs, and the non-adiabatic couplings are dominated by two states with adiabatic energies 
\begin{align}
E_{1}(t)=\pm\Omega_{0}\sqrt{\gamma_{m}^{N}(t)}.
\end{align}
\item
\emph{Large single-photon detuning}: If $\Delta\gg\Omega_{0}$, the adiabatic energy is approximately 
\begin{align}
E_{1}(t)\approx-\frac{\Omega_{0}^{2}}{\Delta}\gamma_{m}^{N}(t),
\end{align}
which is smaller than in the resonant case. Larger values of $\Delta$ require a higher Rabi frequency or a slower variation of $\widehat{H}_{I}(t)$ in order to ensure adiabatic following of the state $\vert\varphi_{0}(t)\rangle$. 
\end{itemize}
In order to minimise the various types of decoherence, it is in general desirable to use pulses with the shortest possible duration. In addition, in order to suppress the effects of spontaneous emission in this many-particle STIRAP, the process must also remain adiabatic.
Increasing the value of $\Omega_{0}$ permits the use of shorter pulses while retaining a high level of adiabaticity. The maximum value of $\Omega_{0}$ is restricted by the condition that $\Omega_{0}$ must remain significantly less than the trap frequency in order to satisfy the rotating-wave approximation and to prevent unwanted excitations in off-resonant vibrational sidebands [see Eq. (\ref{james_condition}) and Ref. \cite{james1998}]. A second route to achieving shorter interaction steps whilst remaining adiabatic is to use resonant pulses, with $\Delta=0$. Use of resonant Raman coupling represents another significant advantage of utilising a dark state, since by being much faster than traditional (off-resonant) Raman-coupled schemes, the process has a natural insensitivity to heating effects.
A contour plot which demonstrates the dependence of the fidelity on $\Omega_{0}$ and $\Delta$ [in the presence of heavy losses from $\vert{}e\rangle$] is shown in Fig. \ref{contour_plots}.

%%%%%%%%%%%%%%%%%%%%%%%%%%%%%%%%%%%%%%%%%%%%%%%%%%%%%%%%%%%%%%%%%%%%%%%%%%%%%%%%%%%%%%%%%%%%%%%%%%%%%%
\subsubsection{Dependence on $m$ and $N$} 
\label{m_and_N}
%%%%%%%%%%%%%%%%%%%%%%%%%%%%%%%%%%%%%%%%%%%%%%%%%%%%%%%%%%%%%%%%%%%%%%%%%%%%%%%%%%%%%%%%%%%%%%%%%%%%%%
%
In general, $\gamma_{m}^{N}(t)$ is a complicated function of both the number of excitations and the number of ions. A numerical evaluation shows that the area \mbox{$\half\Omega_{0}\int_{-\infty}^{\infty}\gamma_{m}^{N}(t')\;dt'$} 
decreases weakly with $m$ and increases in proportion to $\sqrt{N}$ for large $N$. However, we note that a characteristic feature of adiabatic passage in chainwise-coupled systems is that the departure from adiabaticity is determined predominantly by the weakest coupling in the chain \cite{vitanov1998}. For the current system, this is always \mbox{$\lambda_{b,0}(t)=\Omega_{b}(t)/2$}, which is independent of both $m$ and $N$. Indeed, an analytic expansion of $\gamma_{m}^{N}(t)$ for resonant pulses, large $N$, and  for the first few values of $m$ shows that:
\begin{subequations}
\begin{align}
\tau< 0 & 
\begin{cases}
\vert{}E_{1}(t\rightarrow-\infty)\vert\longrightarrow\frac{1}{2}\Omega_{b}(t), 
\\
\vert{}E_{1}(t\rightarrow\infty)\vert\longrightarrow\frac{1}{2}\Omega_{a}(t)\sqrt{N}, 
\end{cases}
\\
\tau>0 & 
\begin{cases}
\vert{}E_{1}(t\rightarrow-\infty)\vert\longrightarrow\frac{1}{2}\Omega_{a}(t)\sqrt{N}, 
\\
\vert{}E_{1}(t\rightarrow\infty)\vert\longrightarrow\frac{1}{2}\Omega_{b}(t). 
\end{cases}
\end{align}
\end{subequations}
Therefore, for both forward and reverse \mbox{STIRAP}, non-adiabatic couplings are at their most potent during the tail of the pulse $\Omega_{b}(t)$. Since the adiabatic energy $E_{1}(t)$ is independent of both $m$ and $N$ in this region, we conclude that the adiabaticity condition for Dicke state preparation does not depend strongly on either $m$ or $N$.
%

%%%%%%%%%%%%%%%%%%%%%%%%%%%%%%%%%%%%%%%%%%%%%%%%%%%%%%%%%%%%%%%%%%%%%%%%%%%%%%%%%%%%%%%%%%%%%%%%%%%%%%
\subsubsection{Dependence on $\Gamma$}
\label{Gamma_dependence}
%%%%%%%%%%%%%%%%%%%%%%%%%%%%%%%%%%%%%%%%%%%%%%%%%%%%%%%%%%%%%%%%%%%%%%%%%%%%%%%%%%%%%%%%%%%%%%%%%%%%%%
%
It is well-known that the \mbox{STIRAP} process is robust to moderate losses, since the adiabatic transfer state $\vert\varphi_{0}(t)\rangle$ does not feature the intermediate state, $\vert{}e\rangle$.
However, for very large decay rates, spontaneous emission can deteriorate the fidelity of the population transfer. This deterioration takes place through two physically distinct routes:
(i) In the absence of losses, it is possible for population to leave the dark state temporarily via the non-adiabatic couplings given in Eq. (\ref{nonadiabatic_couplings}) but to return to $\vert\varphi_{0}(t)\rangle$ by the end of the process \footnote{This population return derives from a higher-order adiabatic process that can be understood by means of the super-adiabatic approach of Berry \cite{berry1990, lim1991}.}. However, for $\Gamma\neq0$, population which leaves $\vert\varphi_{0}(t)\rangle$ may be irretrievably lost. (ii) The strength of the non-adiabatic couplings is dependent on $\Gamma$, and so increasing the rate of spontaneous emission can alter the requirements for adiabatic following. In the extreme case, where $\Omega_{0}/\Gamma\rightarrow0$, this can lead to quantum overdamping, wherein the system becomes frozen in its initial state \cite{vitanov1997c}.

A thorough analysis of the effects of losses in three-state \mbox{STIRAP} (which corresponds to a single excitation in this model) has been performed for the cases of pure dephasing \cite{ivanov2004}, decay into the computational basis states, $\vert0\rangle$ and $\vert1\rangle$ \cite{ivanov2005}, and decay into other states \cite{vitanov1997c}.
When the decay from $\vert{}e\rangle$ is into the computational basis only, the second loss mechanism given above is dominant and the transfer efficiency is found to be \cite{ivanov2005}
\begin{align}
\textrm{Transfer efficiency} = 1 - \exp\left[-\sqrt{\frac{\pi}{2}}\left(\frac{\Omega_{0}^{2}T}{\Gamma}\right)\right].
\label{Gamma_condition}
\end{align}
Conversely, if the state $\vert{}e\rangle$ relaxes predominantly into states outside of the computational basis, then the first relaxation channel given above can also play an important role, and the effects of losses are more pronounced. Nonetheless, the relevant parameter for adiabaticity is still $\Omega_{0}^{2}T/\Gamma$. A quantitative assessment of the effects of spontaneous emission expected with current experimental parameters is given in section \ref{spontaneous_emission}.

%================================================================================================================================================================
\begin{figure}
\includegraphics[width=0.9\columnwidth]{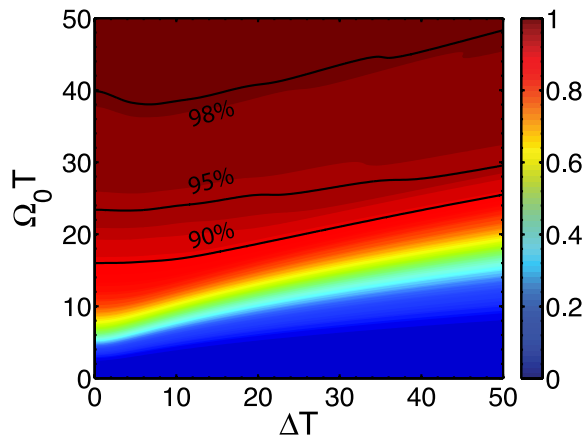}
\caption{Fidelity of the Dicke state preparation as a function of the Rabi frequency, $\Omega_{0}$, and single-photon detuning, $\Delta$, for $m=2, N=5, \tau=-0.6T$ and in the presence of heavy losses ($\Gamma{}T=2$). These are the same parameters as in Fig. \ref{dark_fig_all}. The large region of high fidelity at large $\Omega_{0}T$ and small $\Delta{}T$ show that the procedure is robust to significant fluctuations in the experimental parameters.}
\label{contour_plots}
\end{figure}
%================================================================================================================================================================

%%%%%%%%%%%%%%%%%%%%%%%%%%%%%%%%%%%%%%%%%%%%%%%%%%%%%%%%%%%%%%%%%%%%%%%%%%%%%%%%%%%%%%%%%%%%%%%%%%%%%%
\subsection{Dependence on pulse delay}
\label{pulse_delay}
%%%%%%%%%%%%%%%%%%%%%%%%%%%%%%%%%%%%%%%%%%%%%%%%%%%%%%%%%%%%%%%%%%%%%%%%%%%%%%%%%%%%%%%%%%%%%%%%%%%%%%
%
From the results shown in subsections \ref{example_case} and \ref{adiabaticity}, it is clear that for a counter-intuitive pulse ordering, the dark state $\vert\varphi_{0}(t)\rangle$ connects the initial and target states, following the pathway described in Sec. \ref{generation}. In this case, allowing a significant decay rate has almost no effect on the dynamics, providing that the evolution remains adiabatic, since $\vert\varphi_{0}(t)\rangle$ is a non-decaying state. For an intuitive pulse ordering, the situation is more complicated; it is possible that some population is transferred to the target state, via another, non-dark adiabatic state \cite{vitanov1997b}. In this scenario, admitting a finite decay rate, $\Gamma$, should greatly reduce the overall fidelity, since the intermediate states are then able to decay.

\begin{figure}
\includegraphics[width=0.65\columnwidth]{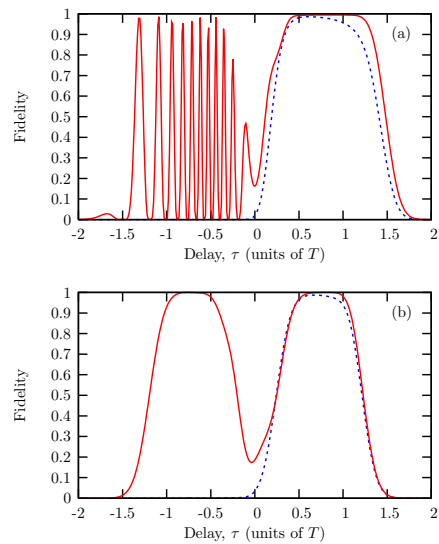}
\caption{Fidelity of the transfer process vs. the time delay, $\tau$ between the pulses.
The solid red curves show the idealised dynamics in the absence of spontaneous emission, while the blue dashed curves include decay from the upper level at rate $\Gamma$. Parameters match those in Fig. \ref{dark_fig_all}, i.e. $m=2, N=5$, $\Omega_{0}T=50, \Gamma{}T=2$ and (a) $\Delta{}T=0$; (b) $\Delta{}T=30$.} 
\label{delay_fig}
\end{figure}

In order to test the above reasoning, the fidelity is plotted against pulse delay $\tau$ in Fig. \ref{delay_fig} for forward-\mbox{STIRAP}. Figure \ref{delay_fig}a shows the case where the single-photon detuning, $\Delta$ is zero, and Fig. \ref{delay_fig}b shows an off-resonant case, with $\Delta{}T=30$.
Positive $\tau$ corresponds to a counter-intuitive pulse ordering, and in this situation, the transfer process can tolerate significant variations in $\tau$, while retaining a transfer efficiency close to $100\%$. Note that this robustness is more pronounced in the resonant case.
Also, allowing a significant decay rate from the state $\vert{}e\rangle$ has a negligible effect on the dynamics, which explains the close overlap between the solid and dashed curves in Fig. \ref{delay_fig} for positive $\tau$. 

For intuitively-ordered pulses, the transfer efficiencies can still be high for certain values of the pulse delay in the idealised case of $\Gamma{}=0$. However, the system then evolves though an adiabatic state which has the potential to decay, and therefore the fidelity is very low when a significant decay rate is introduced. This explains the behaviour of the blue dashed curve for negative $\tau$.

%================================================================================================================================================================
%================================================================================================================================================================
%

\subsubsection{Adiabatic elimination of the state $\vert{}e\rangle$}
\begin{figure}
\includegraphics[width=0.75\columnwidth]{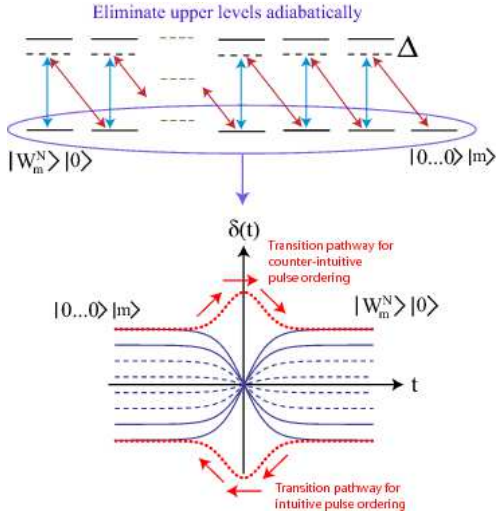}
\caption{For a large single-photon detuning, $\Delta$, it is possible to eliminate the $\epsilon=1$ manifold adiabatically. In this case, the chain-STIRAP model described by Eq. (\ref{matrix_form}) reduces to an $(m+1)$-level ``bow-tie'' energy pattern. The slope of the effective detuning, $\delta(t)$, of levels in the reduced state space is controlled by the pulse delay, $\tau$. It is known that in a bow-tie crossing, transfer efficiency does not depend on the sign of the chirp rate, and this explains the symmetry of the red curve in \mbox{Fig. \ref{delay_fig}b} with respect to  the sign of $\tau$. Including a finite decay rate breaks this symmetry.} 
\label{adiabatic_elimination_fig}
\end{figure}

It is interesting to note that the red curve in \mbox{Fig. \ref{delay_fig}b} is nearly symmetric with respect to the pulse delay. This can be understood through an adiabatic elimination of the states in the $\epsilon=1$ manifold, which is appropriate when the single-photon detuning $\Delta$ is large. In this case, the chain-STIRAP structure given in Eq. (\ref{matrix_form}) reduces to an $(m+1)$-level ``bow-tie'' coupling, as outlined in Fig. \ref{adiabatic_elimination_fig}. Different pulse delays in the original picture are mapped onto different \emph{chirp rates} in the bow-tie picture. [The Gaussian time-dependence given by Eq. (\ref{pulses}) results in an approximate sech-tanh dependence for the bow-tie couplings and detunings \cite{vitanov1997c} -- i.e. a multi-level Allen-Eberly model \cite{allen1987}.]
Such a bow-tie coupling scheme was discussed in \cite{Dicke_bow_tie_2008}, and arises naturally in the preparation of Dicke states using two-level ions and a frequency-chirped laser pulse. As shown in \cite{Dicke_bow_tie_2008}, and sketched in Fig. \ref{adiabatic_elimination_fig}, an adiabatic pathway connecting the initial and target states always exists in the bow-tie model, regardless of the sign of the chirp rate. This explains why the red curve in Fig \ref{delay_fig}b is symmetric under inversion of the sign of $\tau$. 
Admitting a finite decay rate breaks this symmetry, since, for negative $\tau$, the adiabatic state used for the state preparation is a \emph{decaying state} whereas for positive $\tau$ this adiabatic state is non-decaying. 
The above behaviour is closely analogous to that found in traditional STIRAP for the case of three levels \cite{vitanov1997b}, here observed in a many-particle, multi-level scenario.
As such, it provides strong confirmation that the dynamics do in fact take place via a multi-particle dark state, as claimed above.

%================================================================================================================================================================
%================================================================================================================================================================
%

%%%%%%%%%%%%%%%%%%%%%%%%%%%%%%%%%%%%%%%%%%%%%%%%%%%%%%%%%%%%%%%%%%%%%%%%%%%%%%%%%%%%%%%%%%%%%%%%%%%%%%
\section{Technical considerations}
\label{technical}
%%%%%%%%%%%%%%%%%%%%%%%%%%%%%%%%%%%%%%%%%%%%%%%%%%%%%%%%%%%%%%%%%%%%%%%%%%%%%%%%%%%%%%%%%%%%%%%%%%%%%%
%
Below, we assess the effects of decoherence for currently achievable experimental parameters.
We note that the effects of spontaneous emission can be reduced by using longer pulses (as this increases the adiabaticity of the process), while vibrational heating is reduced by using shorter pulses. Therefore, in order to maximise the overall fidelity, the pulse duration must be chosen in order to achieve a balance between these two mechanisms for decoherence. For concreteness, all values quoted below refer to the specific example of trapped $\textrm{Ca}^{+}$ ions.

%%%%%%%%%%%%%%%%%%%%%%%%%%%%%%%%%%%%%%%%%%%%%%%%%%%%%%%%%%%%%%%%%%%%%%%%%%%%%%%%%%%%%%%%%%%%%%%%%%%%%%
\subsection{Spontaneous emission}
\label{spontaneous_emission}
%%%%%%%%%%%%%%%%%%%%%%%%%%%%%%%%%%%%%%%%%%%%%%%%%%%%%%%%%%%%%%%%%%%%%%%%%%%%%%%%%%%%%%%%%%%%%%%%%%%%%%
%
In order to limit the effects of spontaneous emission, the system must remain in the dark state, $\vert\varphi_{0}(t)\rangle$, at all times. For all of the ion species favoured in contemporary quantum information processing experiments using Raman-coupled qubits, this is a non-trivial requirement, since the decay rate $\Gamma$ greatly exceeds the allowed values of the Rabi frequencies, $\Omega_{a}(t)$ and $\Omega_{b}(t)$. Therefore, the adiabaticity of the state preparation is limited by the effects of overdamping \cite{shore2006}, as described in section \ref{Gamma_dependence}. Equation (\ref{Gamma_condition}) shows that in order to achieve a transfer efficiency of $1-x$ (with $x\ll1$), the minimum pulse timescale  is:
\begin{align}
T\gtrsim\sqrt{\frac{2}{\pi}}\left(\frac{\Gamma{}\ln(1/x)}{\Omega_{0}^{2}}\right).
\label{pulse_timescale}
\end{align}
For ${}^{43}$Ca$^{+}$, the decay rate is \mbox{$\Gamma/2\pi\sim22$MHz} \cite{wineland2003}, while the maximum permitted value of $\Omega_{0}$ is limited by the dual requirements that the vibrational rotating-wave approximation should remain valid \cite{lizuain2008} and that unwanted excitations are not created in vibrational modes other than the centre-of-mass mode \cite{james1998}. A conservative estimate for $\Omega_{0}$ which satisfies both of these requirements is \mbox{$\Omega_{0}\lesssim\nu/10$} \cite{james1998}. Taking a trap-frequency of \mbox{$\nu/2\pi\sim4$MHz} \cite{schmidt2003}, then gives \mbox{$T\sim80\mu$s} in order to achieve a transfer efficiency of $99\%$. 

%%%%%%%%%%%%%%%%%%%%%%%%%%%%%%%%%%%%%%%%%%%%%%%%%%%%%%%%%%%%%%%%%%%%%%%%%%%%%%%%%%%%%%%%%%%%%%%%%%%%%%
\subsection{Heating effects}
\label{heating}
%%%%%%%%%%%%%%%%%%%%%%%%%%%%%%%%%%%%%%%%%%%%%%%%%%%%%%%%%%%%%%%%%%%%%%%%%%%%%%%%%%%%%%%%%%%%%%%%%%%%%%
%
In order to limit the effects of heating, it is desirable to carry out the state preparation as fast as possible, and in this sense the fact that our technique requires only two interaction steps represents a significant advantage. Ultimately, however, Eq. (\ref{pulse_timescale}) sets a lower limit to the duration of each interaction step.

In order to estimate the effects of motional heating we assume that the heating rate per ion is $\sim5$Hz, as recently measured in Ref. \cite{schmidt2003}. Although this value was measured for the isotope ${}^{40}$Ca$^{+}$, we note that the same value is expected for ${}^{43}$Ca$^{+}$ \cite{lucas2007}. Allowing an overall time of $T_{\textrm{Dicke}}\sim6T$ for the two pairs of pulses, we therefore predict the total number of heating events during the state preparation to be on the level of \mbox{$2.4\times10^{-2}$} for a chain of $10$ ions: i.e. an overall fidelity approaching $98\%$.

Finally, we note that the effects of heating can be reduced by several orders of magnitude if another vibrational mode is used, rather than the centre-of-mass mode. In this case, individual addressing of the ions is necessary, with appropriately tuned couplings, in order to realise the equal-coupling Hamiltonian (\ref{Hamiltonian_no_phases}). An auxiliary ion may be incorporated in the chain for the purposes of continuous sympathetic cooling of the centre-of-mass mode \cite{kielpinski2000}. Because the heating rates for higher-order modes are considerably lower than for the centre-of-mass vibrational mode, the effects of motional heating may be greatly suppressed in this case. Use of higher-order modes therefore permits an increase in the pulse timescale $T$ which helps adiabaticity and hence further reduces the effects of spontaneous emission.

%
%%%%%%%%%%%%%%%%%%%%%%%%%%%%%%%%%%%%%%%%%%%%%%%%%%%%%%%%%%%%%%%%%%%%%%%%%%%%%%%%%%%%%%%%%%%%%%%%%%%%%%
\subsection{Uneven spacing of the ions}
%%%%%%%%%%%%%%%%%%%%%%%%%%%%%%%%%%%%%%%%%%%%%%%%%%%%%%%%%%%%%%%%%%%%%%%%%%%%%%%%%%%%%%%%%%%%%%%%%%%%%%
%
In the main body of this article, we have considered a Hamiltonian, $\widehat{H}_{I}(t)$ [given in Eq. (\ref{Hamiltonian_no_phases})], which is symmetric under interchange of any two ions. Now, as discussed in subsection \ref{definition}, the Hamiltonian for a chain of trapped ions coupled by a pair of laser pulses is actually
$\tilde{H}_{I}(t)$ [given in Eq. (\ref{Hamiltonian})], which is related to $\widehat{H}_{I}(t)$ via the phase transformation, (\ref{phase_transformation}). The solution, $\vert\tilde{\psi}(t)\rangle$ to the Schr\"odinger equation using $\tilde{H}_{I}(t)$ is therefore related to $\vert\psi(t)\rangle$ given above by:
\begin{align}
\vert\tilde{\psi}(t)\rangle = \mathbf{U}\left\vert\psi(t)\right\rangle,
\end{align}
and so, with the Hamiltonian given in equation (\ref{Hamiltonian}), the two preparation steps described in equations (\ref{stage_1}) and (\ref{stage_2}) are simply replaced by
\begin{align}
\exp\bigg[i\sum_{j=1}^{m}(\phi_{j}^{b}-\phi_{j}^{a})\bigg]\vert\underbrace{1\ldots1}_{m}\underbrace{0\ldots0}_{N-m}\rangle\longrightarrow\vert\underbrace{0\ldots0}_{N}\rangle\vert{}m\rangle.
\end{align}
and
\begin{align}
\vert0\ldots0\rangle\vert{}m\rangle\longrightarrow\exp\bigg[i\sum_{j=1}^{N}(\phi^{b}_{j}-\phi^{a}_{j})\vert{}1\rangle_{j}\langle{}1\vert_{j}\bigg]\vert{}W_{m}^{N}\rangle\vert0\rangle.
\end{align}
respectively. Physically, these unwanted phase terms arise due to the uneven spacing of the ions along the trap axis. They can be corrected by a controlled rotation of the state of each ion, $j$, so that $\vert{}1\rangle_{j}\longrightarrow{}e^{i(\phi_{j}^{a}-\phi_{j}^{b})}\vert{}1\rangle_{j}$.
without affecting $\vert{}0\rangle_{j}$ or $\vert{}e\rangle_{j}$. Alternatively, these constant phases may be absorbed into the definition of the computational basis states. We emphasise that $\phi_{j}^{a,b}$ are \emph{known, constant phases}, which are unchanged by fluctuations in the interaction parameters, and that this is greatly preferable to the dynamical phases arising in many other approaches.

%%%%%%%%%%%%%%%%%%%%%%%%%%%%%%%%%%%%%%%%%%%%%%%%%%%%%%%%%%%%%%%%%%%%%%%%%%%%%%%%%%%%%%%%%%%%%%%%%%%%%%
\subsection{Spatial profile of laser beams and parameter fluctuations}
%%%%%%%%%%%%%%%%%%%%%%%%%%%%%%%%%%%%%%%%%%%%%%%%%%%%%%%%%%%%%%%%%%%%%%%%%%%%%%%%%%%%%%%%%%%%%%%%%%%%%%
%
The procedure described above can tolerate significant variations in the experimental parameters. Figure \ref{contour_plots} makes it clear that the technique is robust against sizeable fluctuations in $\Omega_{0}$ and $\Delta$, and Fig. \ref{delay_fig} shows that imperfections in the pulse timing have almost no effect on the overall fidelity. However, we note that as in previous STIRAP-type models, our technique does require that the Raman-coupled transition between $\vert0\rangle\leftrightarrow\vert{}1\rangle$ satisfies a condition of two-photon resonance, i.e.
\begin{align}
\omega_{a}-\omega_{b}+\nu = \omega_{0e}-\omega_{1e},
\label{two_photon_resonance}
\end{align}
and does not tolerate relative fluctuations in the frequencies $\omega_{a}$ and $\omega_{b}$.

We also note that in a practical setting, some slight variations in the laser intensity at the position of each ion might be expected. In order to quantify the effects of such variations, we have simulated the state preparation using laser pulses with a Gaussian spatial profile and a $10\%$ variation in the pulse intensities from the centre to the edge of the ion chain. In this case, and using the same parameters as in Fig. \ref{dark_fig_all}, the overall fidelity remained above $98\%$. 

Finally we remark on another considerable advantage of our scheme -- namely that the state vector does not acquire any dynamical or geometric phase; the state $\vert\varphi_{0}(t)\rangle$ has zero energy in the interaction picture, and the relative phases of the laser pulses remain constant during the adiabatic passage. The absence of dynamical phases represents a significant advantage of our scheme, since in other approaches, relatively small unknown imperfections in the experimental parameters can lead to sizeable unknown phase factors between different components of the state vector and this can dramatically reduce the overall purity of the state.

%%%%%%%%%%%%%%%%%%%%%%%%%%%%%%%%%%%%%%%%%%%%%%%%%%%%%%%%%%%%%%%%%%%%%%%%%%%%%%%%%%%%%%%%%%%%%%%%%%%%%%
\section{Conclusions}
\label{conclusions}
%%%%%%%%%%%%%%%%%%%%%%%%%%%%%%%%%%%%%%%%%%%%%%%%%%%%%%%%%%%%%%%%%%%%%%%%%%%%%%%%%%%%%%%%%%%%%%%%%%%%%%
%
We have shown that maximally-entangled Dicke states, $\vert{}W_{m}^{N}\rangle$, of arbitrary size may be generated in an ion trap using two pairs of extremely simple laser pulses and remaining in a decoherence-free subspace with respect to spontaneous emission. Starting with $m$ ions in state $\vert{}1\rangle$ and all other ions in state $\vert{}0\rangle$, a pair of overlapping, counter-intuitively ordered pulses -- each of which simultaneously addresses these $m$ ions -- is used to steer the system adiabatically into a phonon
number state with zero ionic excitations and $m$ phonons. 
A second pair of similar pulses -- this time addressing all $N$ ions -- is subsequently used to transfer the system robustly from the $m$-phonon state to the Dicke state, $\vert{}W_{m}^{N}\rangle$.

This extremely simple scenario derives from three theoretical findings: (i) The total number of excitations is preserved by the Hamiltonian, which is also symmetric with respect to interchange of the ions. For an appropriately chosen initial state, the dynamics are thus confined to the subspace of generalised Dicke states [Eq. (\ref{considered_states})], with a given number of excitations; (ii) this symmetric subspace contains a unique multi-ion dark state, in which the upper level $\vert{}e\rangle$ is never populated in any ion. Therefore, providing that the system remains in this dark state, most couplings within the huge symmetric subspace can be neglected and the problem becomes analytically tractable; (iii) remarkably, by applying counter-intuitively ordered overlapping pulses, this dark state can be transformed adiabatically from an initial product state into the Dicke state, $\vert{}W_{m}^{N}\rangle$.

In addition to an extremely simple experimental implementation, our proposal also possesses several other highly attractive features: (i) the dynamical and geometric phase acquired during the whole preparation procedure are both identically zero;
(ii) the proposed technique is adiabatic in nature and hence it is robust against intensity and frequency imperfections; (iii) there is no decoherence arising from spontaneous emission in the adiabatic limit, regardless of the decay rate from the upper level, since the process utilises a dark state. This allows the use of resonant laser pulses which in turn allows shorter pulse durations; (iv) because only two interaction steps are required and these can both be performed very rapidly, heating effects are greatly reduced when compared to other schemes.

\begin{acknowledgments}
We thank P. A. Ivanov for helpful comments and gratefully acknowledge support from the EU ToK project CAMEL, the EU RTN project EMALI, the EU ITN project FASTQUAST and the Bulgarian National Science Fund Grants No. 2501/06 and No. 2517/07.
\end{acknowledgments}

%%%%%%%%%%%%%%%%%%%%%%%%%%%%%%%%%%%%%%%%%%%%%%%%%%%%%%%%%%%%%%%%%%%%%%%%%%%%%%%%%%%%%%%%%%%%%%%%%%%%%%
\appendix
\section{Nonzero eigenvalues of $\widehat{H}_{I}(t)$}
\label{appendix}
%%%%%%%%%%%%%%%%%%%%%%%%%%%%%%%%%%%%%%%%%%%%%%%%%%%%%%%%%%%%%%%%%%%%%%%%%%%%%%%%%%%%%%%%%%%%%%%%%%%%%%
%
In addition to the determinant given in Eq. (\ref{odd_determinant}), we define $M_{2m}$ as follows:
\begin{small}
\begin{align}
M_{2m} \equiv \left\vert
\begin{array}{cccccc}
-E_{\varphi} & \lambda_{b,0} & 0 & \hdots & 0 & 0\\
\lambda_{b,0} & \Delta-E_{\varphi} & \lambda_{a,1} & \hdots & 0 & 0\\
0 & \lambda_{a,1} & -E_{\varphi} &  \hdots & 0 & 0\\
\vdots & \vdots & \vdots &  \ddots & \vdots & \vdots \\
0 & 0 & 0 & \hdots & -E_{\varphi} & \lambda_{b,m-1} \\
0 & 0 & 0 & \hdots  & \lambda_{b,m-1} & \Delta-E_{\varphi}
\end{array}\right\vert.
\label{even_determinant}
\end{align}
\end{small}
The recurrence relations defining $M_{2m+1}$ and $M_{2m}$ are therefore:
\begin{subequations}
\begin{align}
M_{2m+1} & = -E_{\varphi}M_{2m} - \lambda_{a,m}^{2}M_{2m-1},
\\
M_{2m} & = (\Delta-E_{\varphi})M_{2m-1}-\lambda_{b,m-1}^{2}M_{2m-2},
\end{align}
with 
\begin{align}
M_{1} =& -E_{\varphi},
\\
M_{2} = &-E_{\varphi}(\Delta-E_{\varphi})-\lambda_{b,0}^{2},
\end{align}
\end{subequations}
and elimination of all the even-dimensional determinants gives:
\begin{align}
M_{2m+1} = & -\lambda_{a,m}^{2}M_{2m-1}
-E_{\varphi}\Big[(\Delta-E_{\varphi})M_{2m-1}
\nonumber \\
&-\lambda_{b,m-1}^{2}\Big[(\Delta-E_{\varphi})M_{2m-3}
-\lambda_{b,m-2}^{2}\Big[\ldots\Big]\Big]\Big].
\label{long_recurrence}
\end{align}
After recalling that 
\begin{subequations}
\begin{align}
\lambda_{a,\mu} = &\frac{1}{2}\Omega_{0}\sqrt{\mu(N-m+\mu)}\exp\left(\frac{-(t+\tau)^{2}}{T^{2}}\right),
\\
\lambda_{b,\mu} = &\frac{1}{2}\Omega_{0}\sqrt{m-\mu}\exp\left(\frac{-(t-\tau)^{2}}{T^{2}}\right),
\end{align}
\end{subequations}
a close inspection of Eq. (\ref{long_recurrence}) reveals that the $2m$ solutions of Eq. (\ref{characteristic_polynomial}) for which $E_{\varphi}\neq0$ are contained in the roots of an $m^{th}$-order polynomial, $P(z)$, in the dimensionless parameter \mbox{$z\equiv{}E_{\varphi}(\Delta-E_{\varphi})/\Omega_{0}^{2}$}.
Two adiabatic energies, $E_{\varphi}$, can be derived from each root of $P(z)$. 
Importantly for the purposes of subsection \ref{adiabaticity}, the coefficients in $P(z)$ are independent of both $\Omega_{0}$ and $\Delta$. Therefore, the roots of $P(z)$ depend only on the number of excitations, $m$, the number of ions, $N$, and the time, $t$. 
We choose the notation $z=\gamma_{m}^{N}(t)$ for the root of $P(z)$ corresponding to the energy eigenvalue $E_{1}$.
Therefore, the adiabatic energy closest to that of the dark state is:
\begin{align}
E_{1} = & \frac{1}{2}\left(\Delta-\sqrt{\Delta^{2}+4\Omega_{0}^{2}\gamma_{m}^{N}(t)}\right).
\end{align}
%

%%%%%%%%%%%%%%%%%%%%%%%%%%%%%%%%%%%%%%%%%%%%%%%%%%%%%%%%%%%%%%%%%%%%%%%%%%%%%%%%%%%%%%%%%%%%%%%%%%%%%%
% BIBLIOGRAPHY

%\bibliographystyle{apsrev} 
%\bibliography{/Users/Len/Documents/STIRAP_stuff/W_states/W_states_paper/database.bib}

%%%%%%%%%%%%%%%%%%%%%%%%%%%%%%%%%%%%%%%%%%%%%%%%%%%%%%%%%%%%%%%%%%%%%%%%%%%%%%%%%%%%%%%%%%%%%%%%%%%%%%

\end{document}